\documentclass[fleqn,usenatbib]{mnras}

\usepackage{newtxtext,newtxmath}

\usepackage[T1]{fontenc}

\usepackage{epsfig}
\usepackage{graphicx}	
\usepackage{amsmath}	
\usepackage{wasysym}
\usepackage{multirow}
\usepackage{booktabs}
\usepackage{tabularx}
\usepackage{ulem}
\usepackage[dvipsnames]{xcolor}
\usepackage{tablefootnote}

\DeclareRobustCommand{\VAN}[3]{#2}
\let\VANthebibliography\thebibliography
\def\thebibliography{\DeclareRobustCommand{\VAN}[3]{##3}\VANthebibliography}

\newcommand{\kms}{\,km\,s$^{-1}$} 

\newcommand{\Msun}{\ensuremath{\rm M_\odot}}

\newcommand{\Lsun}{\ensuremath{\rm L_\odot}}

\newcommand{\OIII}{[O\,{\sevensize III}]\ }
\newcommand{\OVI}{O\,{\sevensize VI}\ }
\newcommand{\CII}{C\,{\sevensize II}\ }
\newcommand{\CaII}{Ca\,{\sevensize II}\ }
\newcommand{\NaI}{Na\,{\sevensize I}\ }
\newcommand{\CIII}{C\,{\sevensize III}\ }
\newcommand{\CIV}{C\,{\sevensize IV}\ }
\newcommand{\NII}{N\,{\sevensize II}\ }
\newcommand{\NIII}{N\,{\sevensize III}\ }
\newcommand{\NV}{N\,{\sevensize V}\ }

\newcommand{\HeII}{He\,{\sevensize II}\ }
\newcommand{\HeI}{He\,{\sevensize I}\ }

\newcommand{\Jstar}{J0409+3239\,} 

\title[New object with WR phenomenon]{LAMOST J040901.83+329355.6 -- a new Galactic star with Wolf--Rayet characteristics on a post-AGB to CSPN transitional stage}  

\author[Maryeva et al.]{Olga~Maryeva$^{1}$\thanks{E-mail:
olga.maryeva@asu.cas.cz}, Aynur~Abdulkarimova$^{2}$, Sergey~Karpov$^{3}$, Alexei~ Moiseev$^{4,5}$, Dmitry~Oparin$^{4}$ \\ 
\\
$^{1}$ Astronomical Institute of the Czech Academy of Sciences, Fri\v{c}ova 298, 25165 Ond\v{r}ejov, Czech Republic\\
$^{2}$ Shamakhy Astrophysical Observatory, Y.Mammadaliyev, AZ5626, Azerbaijan\\
$^{3}$ Institute of Physics of the  Czech Academy of Sciences, CZ-182 21 Prague 8, Czech Republic  \\
$^{4}$ Special Astrophysical Observatory of the Russian Academy of Sciences, Nizhnii Arkhyz, 369167, Russia\\
$^{5}$ Lomonosov Moscow State University, Sternberg Astronomical Institute,
Universitetsky pr. 13, Moscow 119234, Russia\\
}

\date{Accepted XXX. Received YYY; in original form ZZZ}

\pubyear{2023}

\begin{document}
\label{firstpage}
\pagerange{\pageref{firstpage}--\pageref{lastpage}}
\maketitle

\begin{abstract}

The similarity in physical conditions in winds of low-mass post-asymptotic giant branch stars and evolved massive stars leads to the appearance of an interesting phenomenon of spectral mimicry. Due to that the discovery of every new star with Wolf--Rayet spectrum requires special study of its evolutionary status before it may be included in the list of Galactic Wolf--Rayet (WR) stars. 
A couple of years ago LAMOST J040901.83+323955.6 (hereafter \Jstar) was selected as a WR star in LAMOST spectroscopic database by machine learning methods. In this work we investigated its evolutionary status.
Analyzing the spatial location of \Jstar\, in the Galaxy and its position in the color-magnitude diagram we concluded what \Jstar\, is instead a low mass object with WR phenomenon.  
Its luminosity is $L*=1000~\Lsun$ and effective temperature $T_{\rm eff}$=40,000~K. Using new and archival photometric data we detected irregular variability on time scales from hours to tens of days with amplitude up to $\approx0.2$~mag. Comparison of the spectrum obtained in 2022 with the one from 2014 also shows an evidence of spectral variability. The absence of clearly detected circumstellar nebula does not allow to classify \Jstar\, as  [WR], i.e. a central star of planetary nebula (CSPN). However, its position in Hertzsprung--Russell diagram  suggests that \Jstar\, is a low mass star caught in rare transitional phase to CSPN. Estimation of \Jstar\, mass based on evolutionary tracks shows that it is less than $0.9~\Msun$, and thus the age of the Galaxy is barely enough for the star to evolve to its current stage.

\end{abstract} 




\section{Introduction}\label{intro}

Classical Wolf--Rayet (WR) stars are the final points in the life of massive stars, which massive stars may reach either through the Conti scenario (in the case of single stars, \citet{Conti1975,Conti}),  or through the mass transfer in binary systems \citep{Paczynski1967}. 
WR stars are the stripped cores  of evolved massive stars that lost their external envelopes, have spectra enriched with CNO-cycle products, and burning helium in their cores. Despite being divided into three distinct groups (nitrogen rich WN, carbon rich WC and rare oxygen rich WO), in Hertzsprung--Russell (HR) diagram all WR stars are located in a small area in upper left corner. 

WR stars are known for more than 150 years, and their properties are sufficiently well studied \citep{Crowther2007}.  However there are still open questions about how the formation of WR stars is affected by the metallicity, and the percentage of binary evolution products among them \citep{Neygent2019Galaxies, Shenar2020}. 
Comparison of observed WN/WC ratio with theoretical predictions of single and binary evolutionary scenarios is a good way to test the theory of stellar evolution  \citep{MaederMeynet1994,Massey2015,Rosslowe2015,Neygent2019Galaxies}.

In our Galaxy there are 669 WR stars currently discovered\footnote{Galactic Wolf-Rayet Catalogue accessible at \url{https://pacrowther.staff.shef.ac.uk/WRcat/l}} 
\citep{Rosslowe2015}, while the theory based on current Milky Way star formation rate and duration of WR phase predicts the numbers of WR stars $\sim1200$ \citep{Rosslowe2015}. It means that significant part of WR stars are still not discovered, and the WN/WC ratio may change significantly. 
WR stars in the Galaxy align with spiral arms and the regions of star formation. Thus, the discovery of new WR stars through optical observations has been significantly limited due to the presence of dust extinction. Although the probability of finding a WR star during optical spectroscopic surveys remains \citep{MaizApellaniz2016,WeiZhang}, the majority of discoveries of new WR stars now happens through the observations in infrared (IR) range. For example, \citet{2011Redeyes} identified 60 Galactic WR stars: candidates were selected using the photometry from {\it Spitzer} and Two Micron All Sky Survey (2MASS) databases and confirmed by near-IR spectroscopy. \citet{2012Shara} expanded the list by adding 71 more stars, also preliminary selected from  $J$ and $K$ band 2MASS photometry. 

Moreover, the search for new WR stars is complicated by the effect of spectral mimicry -- in addition to classical WR stars there is also {\it WR phenomenon}, happening when fast-moving, hot plasma is expanding around a hot star \citep{Grafener2011, Vink2015}. The mass loss rate or rather the density of the expanding envelope and the temperature are the critical basic parameters which are responsible for the WR phenomena \citep{vanderHucht1981}, and due to that not only evolved massive stars during WR stage may show it. Besides classical WR stars, WR phenomenon is also observed in young supernovae (SN)  \citet{GalYam2014} and [WR] objects -- central stars of planetary nebulae (CSPNe) \citep{vanderHucht1981, Marcolino2007,Todt2009PhD,Todt2015}, hydrogen deficient post-AGB stars, coming through ``born-again'' evolutionary scenario \citep{Iben1983,WernerHerwig2006} and typically having luminosities less than 25000~$\Lsun$ \citep{Weidmann2020}. The most striking example of the object with WR phenomenon is the famous SS\,433 object -- Galactic microquasar, binary system with black hole candidate \citep{Fabrika2004}, with a Bowen \CIII/\NIII blend which is typical for WR stars clearly visible in the spectrum.

The [WR] objects masquerading  as classical WR stars is a particular case of such mimicry between low mass stars that already passed through asymptotic giant branch (post–AGB), and hypergiants.  
Similarity of wind properties (luminosity $L_*$ to mass $M_*$ ratio of a star, mass loss rate $\dot{M}$) is the reason for spectral mimicry as it was discussed a lot in the works of V.~Klochkova and E.~Chentsov \citep{Klochkova1997,Klochkova2007,Klochkova2014}, as well as by \citet{Lamers1998}. 
Thanks to the results of {\it Gaia} mission \citep{Gaia2016} we now have reliable estimations of distances, and it helps to quickly understand the nature of individual objects and separate low-mass evolved stars from massive stars, and, moreover, to find the objects of unusual origin among low mass stars. For example, based on {\it Gaia} distance estimation \citet{Gvaramadze2019Nature} demonstrated that IRAS\,00500+6713 is a product of merging of two white dwarfs, while the star has WO-type spectrum. Reverse example is PMR\,5 star classified as [WN6] by \citet{Morgan2003} that was recently reclassified as a classical WN \citep{todt_2023}. 

The present paper is devoted to study of LAMOST J040901.83+323955.6 (hereafter J0409+3239) which is a star initially classified as WN  by \citet{Skoda2020}. However, close distance  to it and its low brightness at same time suggest that   \Jstar\ is most probably the low mass object with WR phenomenon. The paper is organised as follows. In Section~\ref{sec:history} we present the object of studies, in Sections~\ref{sec:spectroscopy} and~\ref{sec:photometry} we describe the archival and newly acquired spectroscopic and photometric data, as well as variability of  \Jstar. In Section~\ref{sec:discussion}  we consider spatial position of \Jstar\, in the Galaxy and its location in the HR diagram,  discuss its initial mass and current age. Last Section~\ref{sec:conclusion} presents the conclusion.

\begin{table}
\caption{Parameters of J040901.83+323955.6.}
\label{tab:par}
\begin{tabular}{lll}
\hline
\hline
RA (J2000)   &  $04^{h}09^{m}01^{s}.8343$        & \\
Dec (J2000)  & $+32\degr39\arcmin55\arcsec.7627$ & \\
$l$          &  $164.12944\degr$                 & \\
$b$          &  $-13.9698\degr$                  &  \\
\\
dist (pc)    & $2499.47^{+151.41}_{-162.44}$  &{\it Gaia} DR3 [1] \\
\\
$V$ (mag)    &$14.678\pm0.148$      &   APASS-9       [2]          \\
$B$ (mag)    &$15.006\pm0.218$      &   APASS-9     [2]          \\
$B_P-R_P$    &$0.468\pm0.036$       &   {\it Gaia} eDR3 [3]     \\
             &                      &                           \\ 
$E(B-V)$     &$0.226^{0.02}_{-0.01}$& 3D Dust Mapping  [4]      \\    
\\
\hline
\hline
\multicolumn{3}{l}{[1] -- \citet{Bailer-Jones2021}, [2] -- \citet{APASS9},}\\
\multicolumn{3}{l}{[3] -- \citet{GaiaEDR3}, [4] -- \citet{3DDustMap}.}\\
\end{tabular}
\end{table}

\begin{figure}
\centerline{\resizebox*{1.0\columnwidth}{!}{\includegraphics[angle=0]{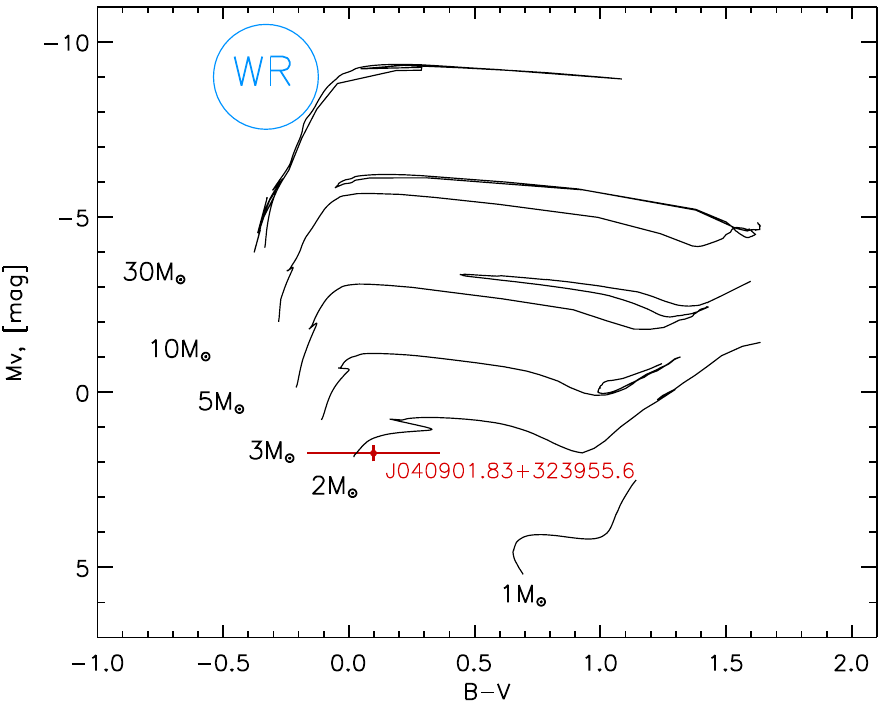}}}
\caption{Position of \Jstar in the color--magnitude diagram. Solid black lines show Geneva  evolutionary tracks \citep{Ekstrom} for solar metallicity and initial rotational velocity 40 per cent of breakup. Position of WR stars is showed schematically.  }
\label{fig:colormag}
\end{figure}

\begin{table*}
\caption{Summary of spectral data used in present work. 
}
\label{tab:log}
\begin{tabular}{lcc ccc c}
\hline 
\multicolumn{1}{c}{Date}     & Telescope                  & Instrument           & Sp. range   & \textbf{Sp. resolution} & Exposure &  Pub.  \\
                             &                            &                      & (\AA)       & \textbf{ ($FWHM$,\AA) }         &  (s)   &     \\
                             
\hline
2014 January 3               & LAMOST                     & low resolution sp.   & 3700-9000   &   $\approx 3.5$  &   4500   &  [1,2]    \\
2022 August 24               & Russian 6-m                & SCORPIO              & 3650-7200   &   $\approx 4.5$        &   3600   &      \\ %
\hline
\multicolumn{7}{l}{[1] -- \citet{Skoda2020}; [2] -- \citet{Sun2021}. }\\  
\end{tabular}
\end{table*}

\begin{figure*}
\centerline{\resizebox*{2.0\columnwidth}{!}{\includegraphics[angle=0]{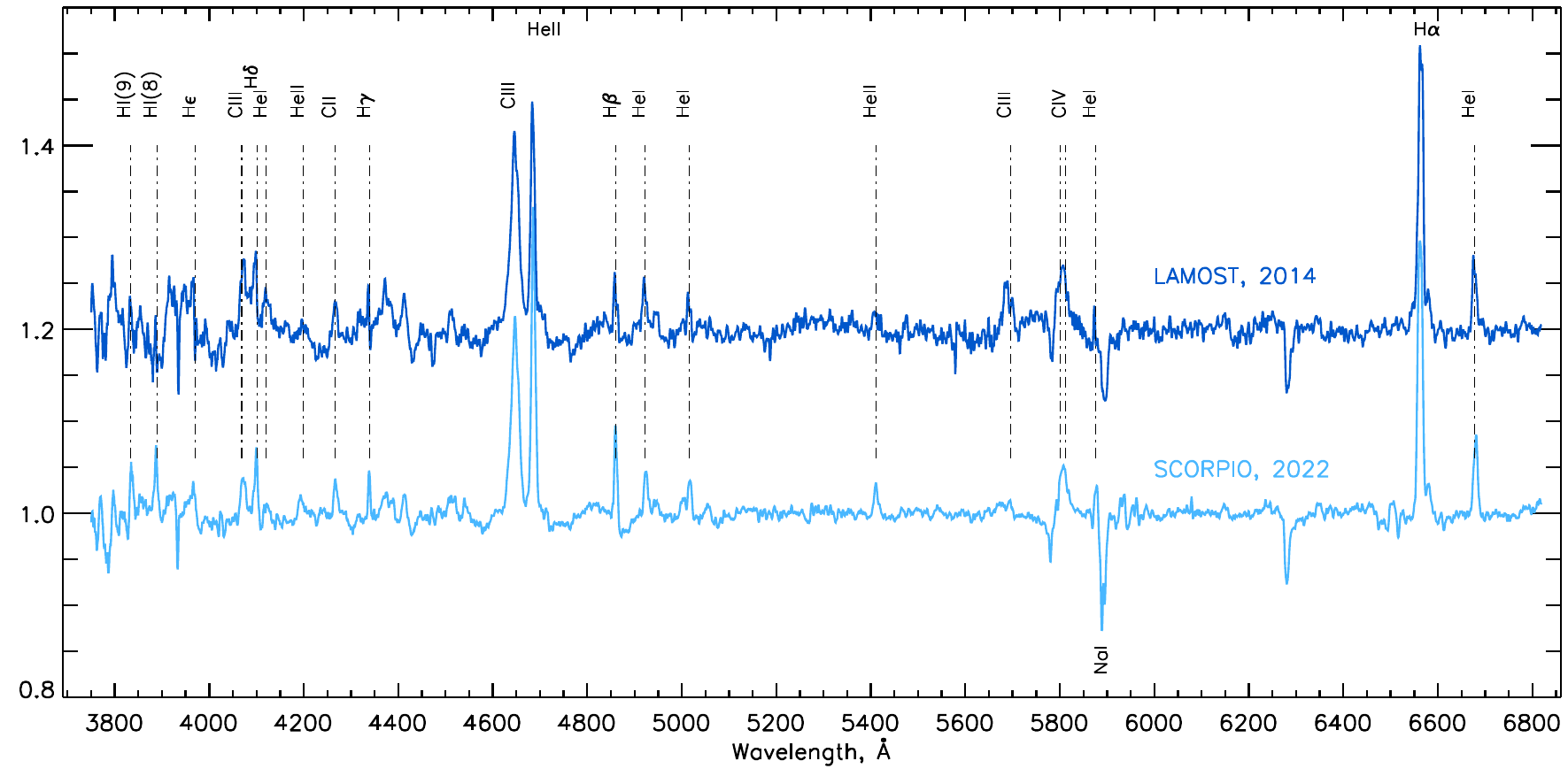}}}
\caption{Continuum normalized spectra of \Jstar obtained in January 2014 and August 2022.  The spectra are vertically shifted for clarity. }
\label{fig:spectrum}
\end{figure*}

\section{History of studies of J040901.83+323955.6}\label{sec:history}

The first spectral observation  of \Jstar\,  was done with Large Sky Area Multi-Object Fiber Spectroscopic Telescope (LAMOST) during LAMOST Spectroscopic Survey of the Galactic Anticentre \citep{LAMOST_Anticentre2015}. \citet{Skoda2020}  selected \Jstar\, in  machine learning based search for emission-line stars in LAMOST archive and classified it as a WN star. Independently \citet{Sun2021} found \Jstar\, in frame of search for new cataclysmic variables (CVs) in LAMOST data and classified it as Nova Like  subtype of CVs. \citet{Sun2021} suggested that \Jstar\, is surrounded by disk, because H$\alpha$ in its spectrum shows  double-peaked profile. \citet{Hou2023} once more mentioned \Jstar\, as a CV without  calculated period. 

The object was also selected in several studies based on photometric data. \citet{Sesar2017RRLyr} added \Jstar\, to the list of RR\,Lyrae stars using machine-learning identification method and multi-epoch,  asynchronous multi-band photometric data from Panoramic Survey Telescope and Rapid Response System (Pan-STARRS).  For \Jstar\, \citet{Sesar2017RRLyr} found the period P=0.2847409137~d.  Moreover, \Jstar\, was included in the first catalog of variable stars of The All-Sky Automated Survey for Supernovae (ASAS-SN) survey \citet{ASAS_firstcatalog}. There the star is mentioned as a non-periodic object with $V$=14.48~mag and amplitude of 0.39~mag.	

Table~\ref{tab:par} presents the coordinates of the object, distance  estimations according to {\it Gaia} third Data Releases (DR3; \citet{Bailer-Jones2021}), photometric data from different catalogs and an estimation of interstellar reddening $E(B-V)$ in the direction of \Jstar. For estimation of $E(B-V)$ we used the distance from {\it Gaia} DR3, and a three-dimensional map of dust reddening\footnote{Interactive viewer is available online at \url{http://argonaut.skymaps.info/}}  \citep{3DDustMap}. Taking into account the distance and  $E(B-V)$ we estimated absolute visual magnitude $M_V$ of \Jstar\, and determined its location in the colour--magnitude diagram (Figure~\ref{fig:colormag}).  Figure~\ref{fig:colormag} clearly shows that \Jstar\, lies in the region of low mass stars  and the object significantly differs from WR stars by magnitude. Exotic products of binary evolution mentioned above like galactic microquasar SS\,433 \citep{Fabrika2004} or two white dwarfs merger product IRAS\,00500+6713 \citep{Gvaramadze2019Nature} are also significantly more luminous. 
Such a large difference in luminosity cannot be explained by interstellar absorption, because the total absorption in the Galaxy in this direction does not exceed 1~mag  \citep{Schlegel1998,Schlafly2011}. 
As we will show in Section~\ref{sec:galaxy}, incorrect determination of the distance also cannot explain it. Therefore we may conclude that \Jstar\, is not a \textit{bona fide} Wolf-Rayet star, but a  low mass object with WR phenomenon.

\section{Spectroscopy}\label{sec:spectroscopy}

New low resolution spectrum of \Jstar\, was obtained in August 24, 2022 with the Russian 6-m telescope with the SCORPIO-2  multimode focal reducer\footnote{SCORPIO is Spectral Camera with Optical Reducer for Photometric and Interferometric Observations.}  \citep{scorpio2} with the VPHG\,1200@540 grism, which provided a spectral range of 3650--7200~\AA{} and a spectral resolution  $FWHM\approx 4.5$\AA{} with 1 arcsec  slit width. New E2V CCD261-84 (2K$\times$4K) was used as a detector \citep{Afanasieva2023}. The spectrophotometric standard star G191B2b was observed for flux-calibration on the same night.  Data reduction was done using IDL-based packages as it was described in our previous papers \citep[e.g.][]{mn112}. Figure~\ref{fig:spectrum} shows two spectra of \Jstar\ --  the new one and previous spectrum obtained with LAMOST on January 3rd, 2014. The joint observational log is given in Table~\ref{tab:log}. 

\begin{figure*}
\centerline{\resizebox*{2.0\columnwidth}{!}{\includegraphics[angle=0]{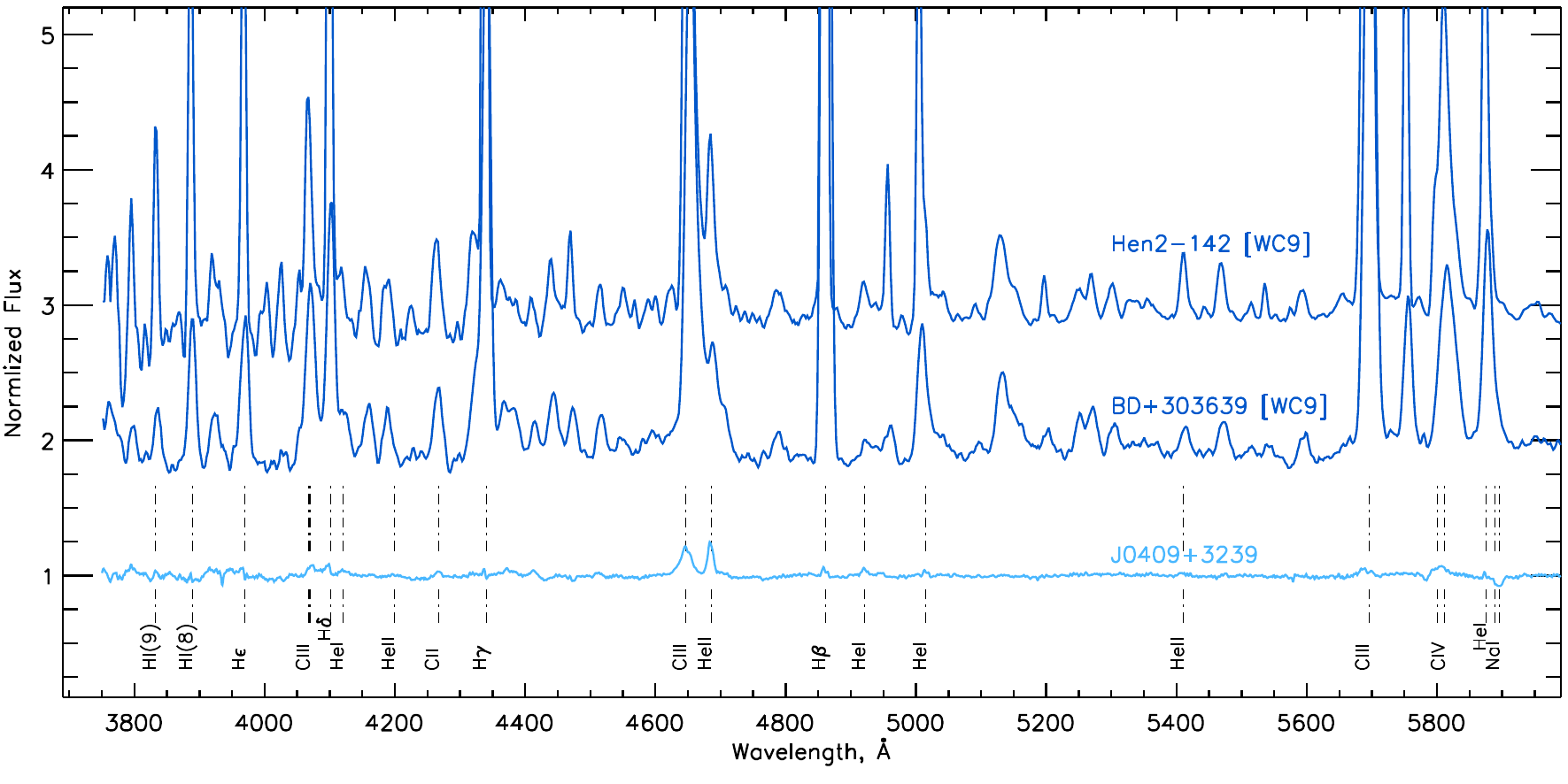}}}
\caption{Comparison of a normalized spectrum of \Jstar, obtained with LAMOST, with ones of [WC9] stars Hen\,2-142 and BD+303639, which are taken from \citet{Hajduk2015}.  The latter spectra are also continuum normalized, and shifted vertically for clarity.}
\label{fig:spectrum_wc}
\end{figure*}

The spectrum  of \Jstar\, contains strong  emissions of  \CII~$\lambda$4267 and \CIII~$\lambda$4647, 4650, 4652 lines. The last one were identified as \NIII by \citet{Skoda2020} and as \CIII+\NIII  by \citet{Sun2021}, however due to the absence of other nitrogen lines we unambiguously identify the bump at 4650~\AA\ as \CIII lines. The \HeII~$\lambda$4686    is the strongest line in the spectrum, it is significantly stronger than another \HeII~$\lambda$5412 line. Also broad emission lines of \CIV~$\lambda$5801, 5812 and weak \HeI are visible.  Presence of helium and carbon lines in the spectrum is a clear evidence for  far evolved stage of the star. Besides carbon and helium, there are also hydrogen lines  of both Balmer and Paschen series in the spectrum. H$\alpha$ has double-peaked profile in the LAMOST spectrum.  
Due to lower resolution of SCORPIO-2 we do not resolve this feature in the 2022 spectrum. We will discuss the nature of hydrogen lines in Section~\ref{sec:discussion}. 
In the spectrum there are also narrow  absorptions of  \CaII~H/K and \NaI~D formed in the interstellar medium (ISM). Comparison of the spectra obtained in 2014 and in 2022 shows a clear difference -- \CIII~$\lambda$5696 line weakened in 2022 in respect to the spectrum of 2014. 



 To better understand the nature of an object  we carried out its spectral classification. As we see that the star is an evolved low-mass object  with WR phenomenon,  for its spectral classification we used the criteria of \citet{Weidmann2020} who suggested the list of key spectral lines (Table~\ref{tab:cspn}) for preliminary determination of sub-type of CSPNe\footnote{In the original work, these criteria are used for the classification of objects selected by the presence of circumstellar nebula, which \Jstar\ lacks. However,
as we are facing the shortcoming of existing classification scheme for it due to its unique evolutionary status (see Section~\ref{sec:discussion}), we decided to still opt for the one designed for CSPNe.}. Since most of the spectral features in the \Jstar's spectrum  are observed in the emission, and there are no oxygen O and nitrogen N lines, we classified  \Jstar\,  as [WC] star. For more accurate classification we used the criteria from \citet{Crowther1998WC} based on equivalent width ratios of \CIV $\lambda$5801, 5812/\CIII$\lambda$5696 and \CIV $\lambda$5801, 5812/\CII$\lambda$4267. According to these equivalent width ratios \Jstar\,  is a [WC8-9]. 

Although according to formal criteria we may classify \Jstar\,  as [WC8-9] stars, its spectrum significantly differs from the spectra of  typical [WC9] stars, for example, Hen\,2-142 and BD+303639 (Figure~\ref{fig:spectrum_wc}). The main difference is in the intensities of emission lines,  that may be interpreted as a significant difference in mass loss rates. Moreover, nebular forbidden lines such as \OIII~$\lambda4959,5007$ and [\NII]~$\lambda$5755 are absent in the \Jstar's spectrum.

\begin{table}
\caption{Key optical lines to spectral classification of CSPNe, according to \citet{Weidmann2020}. The lines are marked as either absent in the spectrum of \Jstar, present in emission, or in a broad emission.}
\label{tab:cspn}
\begin{tabular}{cl |cl}
\hline
\multicolumn{2}{c}{Lines} & \multicolumn{2}{c}{Lines} \\
\hline
  H$\gamma~\lambda4340$   & emission   & \CIV~$\lambda5806$     & broad emission \\
  \HeI~$\lambda4471$      & --     & \CIV~$\lambda4650$     & -- \\
  \CIII~$\lambda4649$     & emission   & \NV~$\lambda4603$      & -- \\
  \CIII~$\lambda5696$     & broad emission  & \NV~$\lambda4945$      & -- \\
  \HeII~$\lambda4686$     & emission   & \OVI~$\lambda3822$     & -- \\
  \HeII~$\lambda5412$     & emission   & \OVI~$\lambda5290$     & -- \\
\hline
\end{tabular}
\end{table}




\begin{figure*}
\centerline{\resizebox*{2.0\columnwidth}{!}{\includegraphics[angle=0]{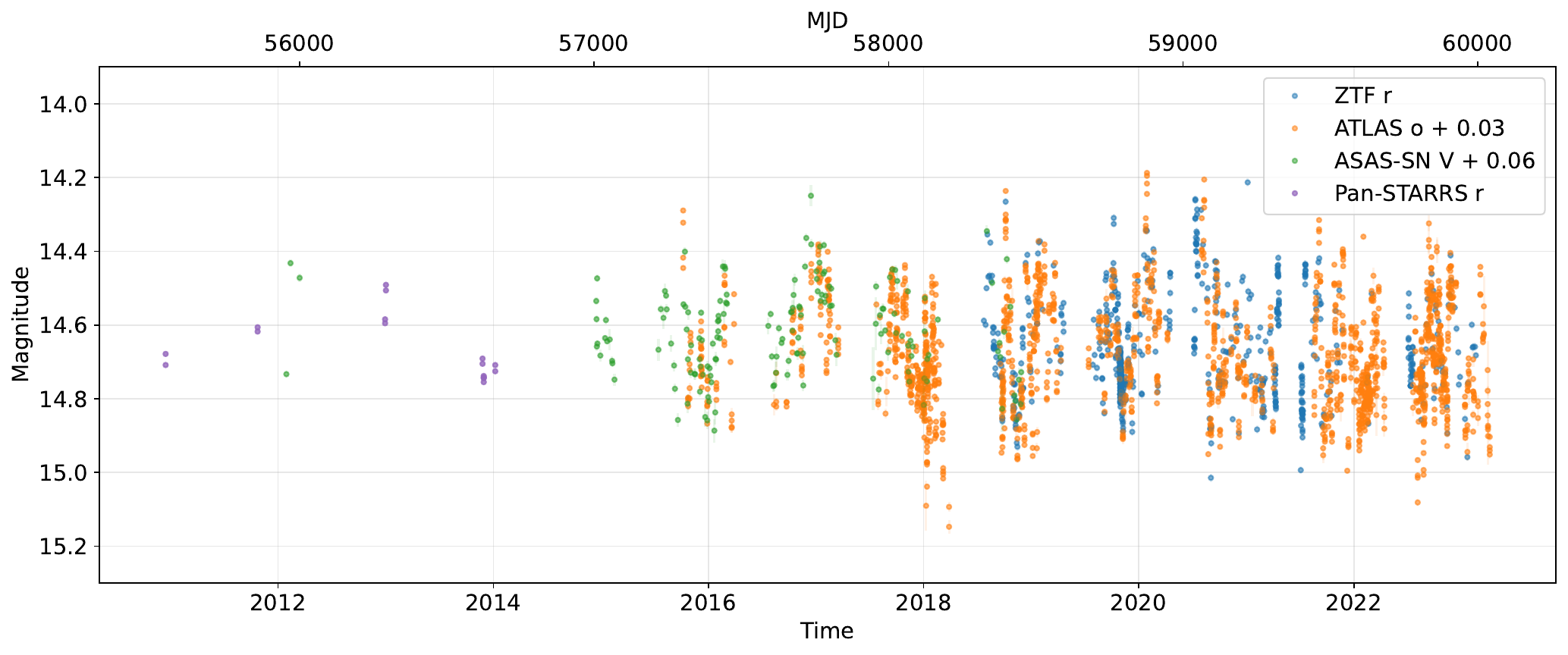}}}
\caption{The light curve of \Jstar over last 12 years using the data from ZTF, ATLAS, ASAS-SN and Pan-STARRS sky surveys.}
\label{fig:lc}
\end{figure*}

\begin{figure*}
\centerline{\resizebox*{2.0\columnwidth}{!}{\includegraphics[angle=0, viewport=0 305 865 590,clip]{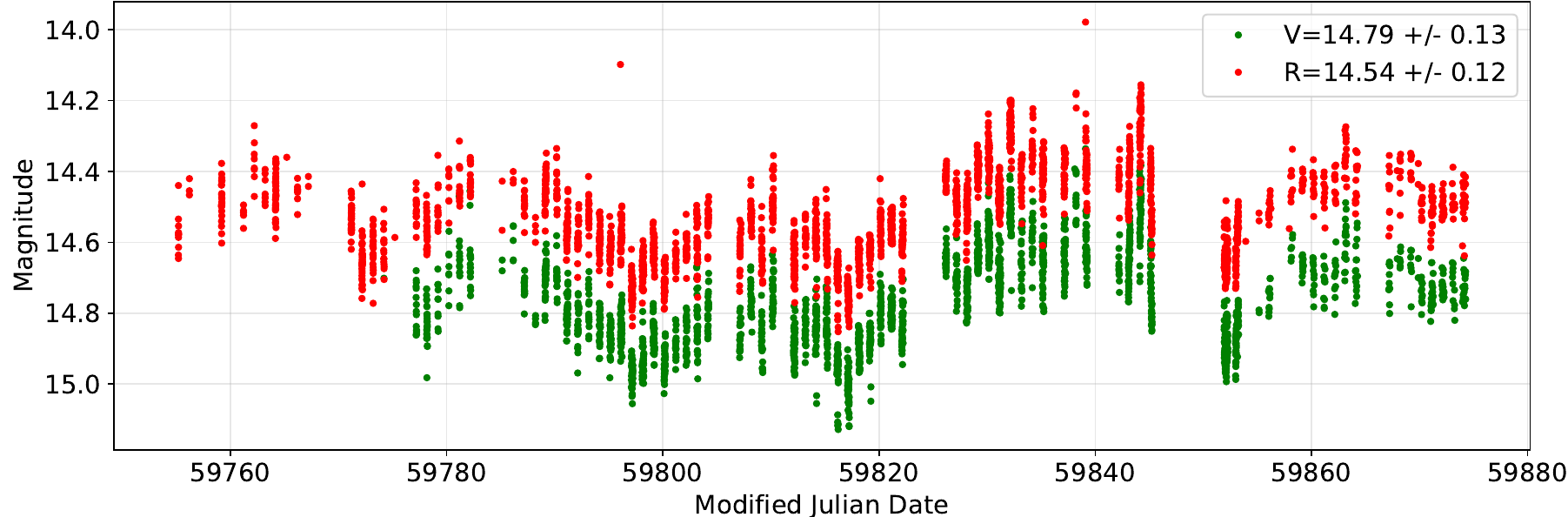}}}
\caption{The light curve of \Jstar based on FRAM-ORM observations (July-August, 2022).}
\label{fig:lc_fram}
\end{figure*}

\begin{figure*}
\centerline{\resizebox*{2.0\columnwidth}{!}{\includegraphics[angle=0]{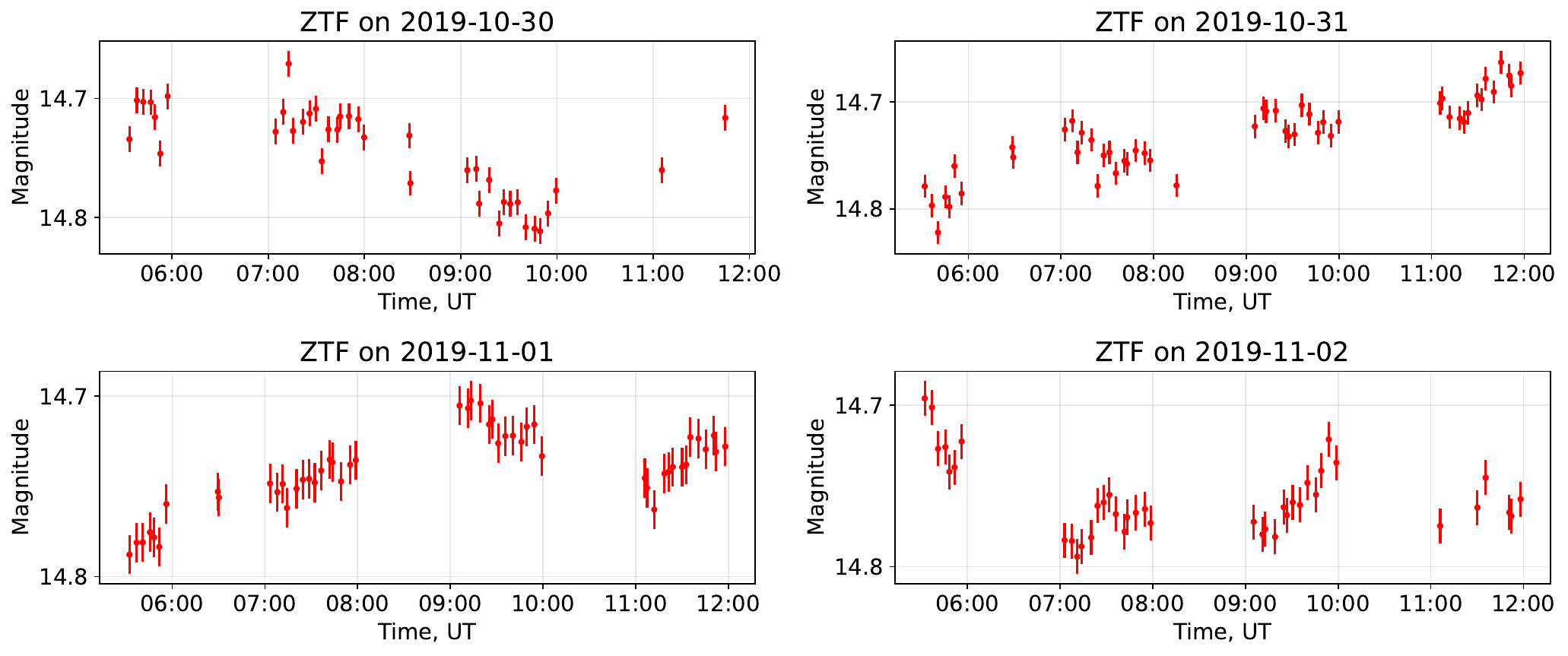}}}

\centerline{\resizebox*{2.0\columnwidth}{!}{\includegraphics[angle=0]{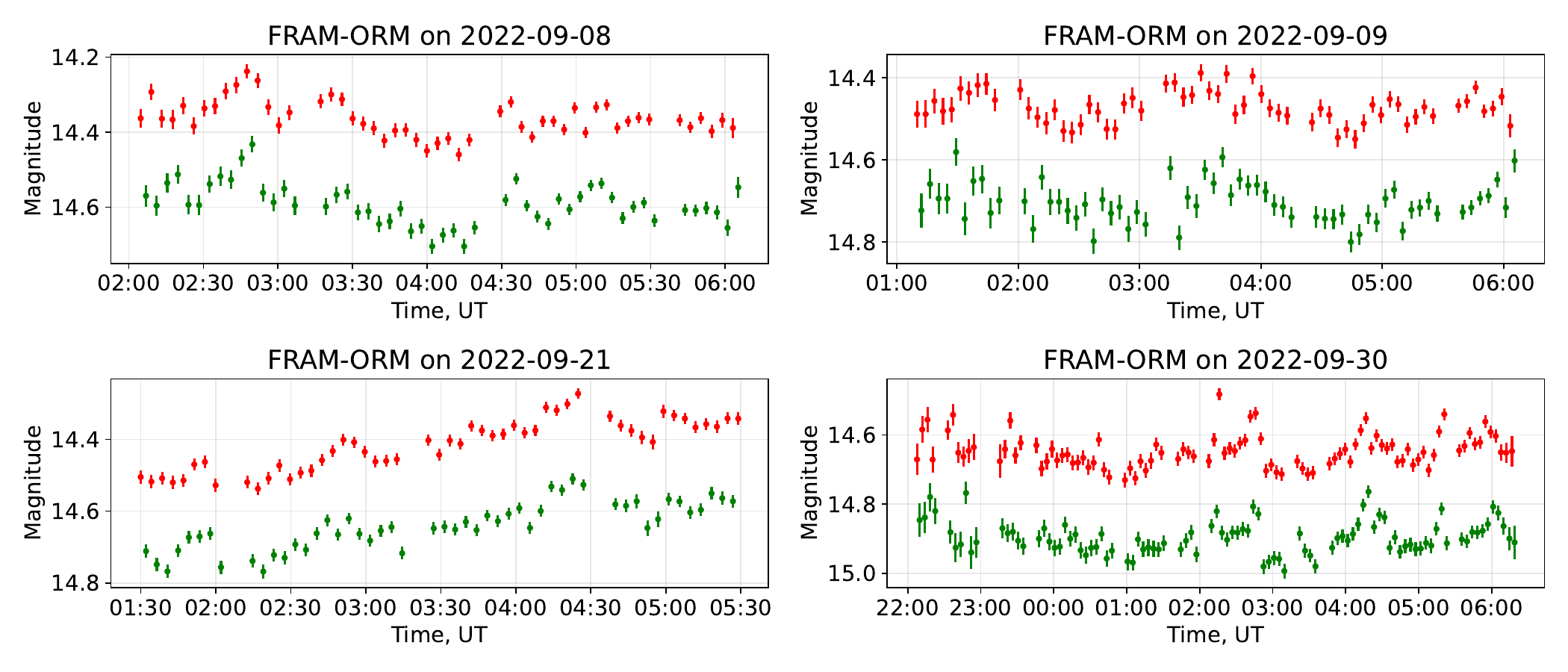}}}
\caption{Intra-night variability of \Jstar\, in ZTF and FRAM-ORM data for four nights from each dataset where densely sampled data were acquired. For ZTF, only $r$ filter data are available, and are shown with red dots. For FRAM-ORM, green dots are measurements in $V$ filter, red -- in $R$. No apparent color variability is seen despite clear variations of intensity in both filters on time scales of tens of minutes only.}
\label{fig:fram_intranight}
\end{figure*}
\section{Photometry}\label{sec:photometry}

In order to assess the photometric variability of \Jstar\,  we utilized the data from Zwicky Transient Facility (ZTF) \citep{ZTF2019Bellm} survey. From measurements collected between March 27, 2018 and February 19, 2023 in ZTF~$g$ and $r$ filters photometric variability of \Jstar is clearly seen. 
We extended the light curve further back in time to better characterize the long-term photometric behaviour of \Jstar\, by combining ZTF data with the measurements from Pan-STARRS1 Survey,  Asteroid Terrestrial-impact Last Alert System (ATLAS) project and ASAS, in the same way as we did it in \citet{mn112}.  Total light curve covering last 12 years is shown in Figure~\ref{fig:lc}. The star does not display any significant change in its photometric behaviour in respect to what we see in ZTF data.

\begin{figure}
\centerline{\resizebox*{\columnwidth}{!}{\includegraphics[angle=0]{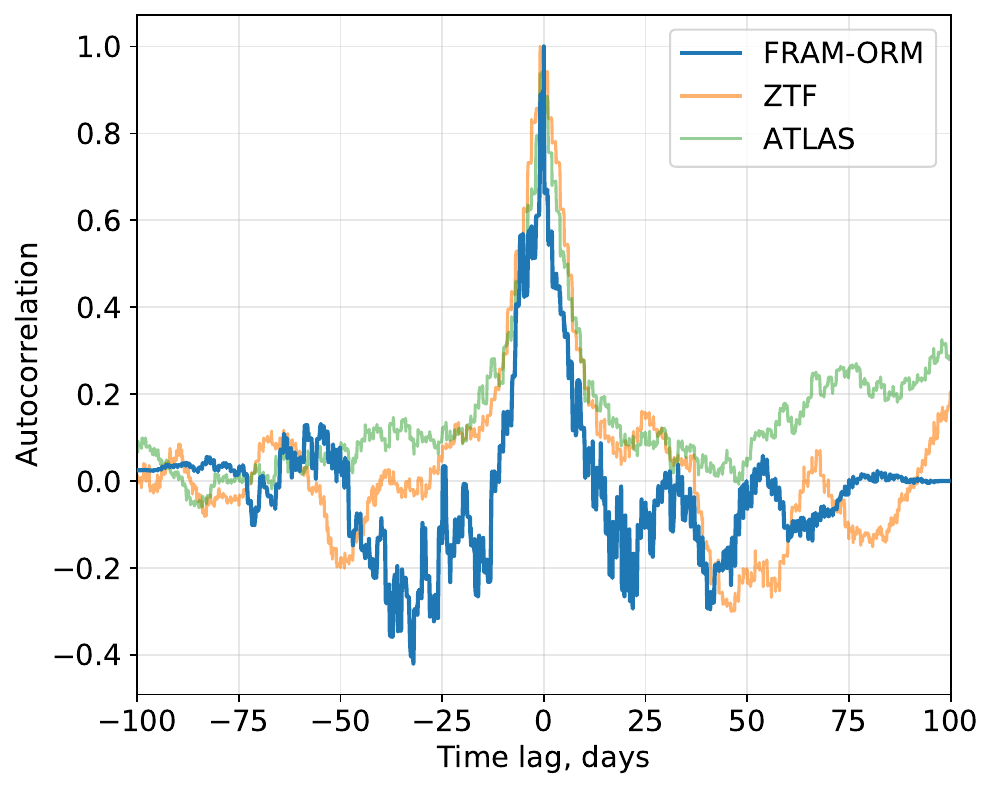}}}
\caption{Autocorrelation function of \Jstar\, lightcurves from different datasets constructed using the method of \citet{sacf} as described in Section~\ref{sec:photometry}. The data from FRAM-ORM is best sampled and most uniform, and thus better captures short-term correlations.}
\label{fig:autocorr}
\end{figure}
\begin{figure}
\centerline{\resizebox*{1.0\columnwidth}{!}{\includegraphics[angle=0]{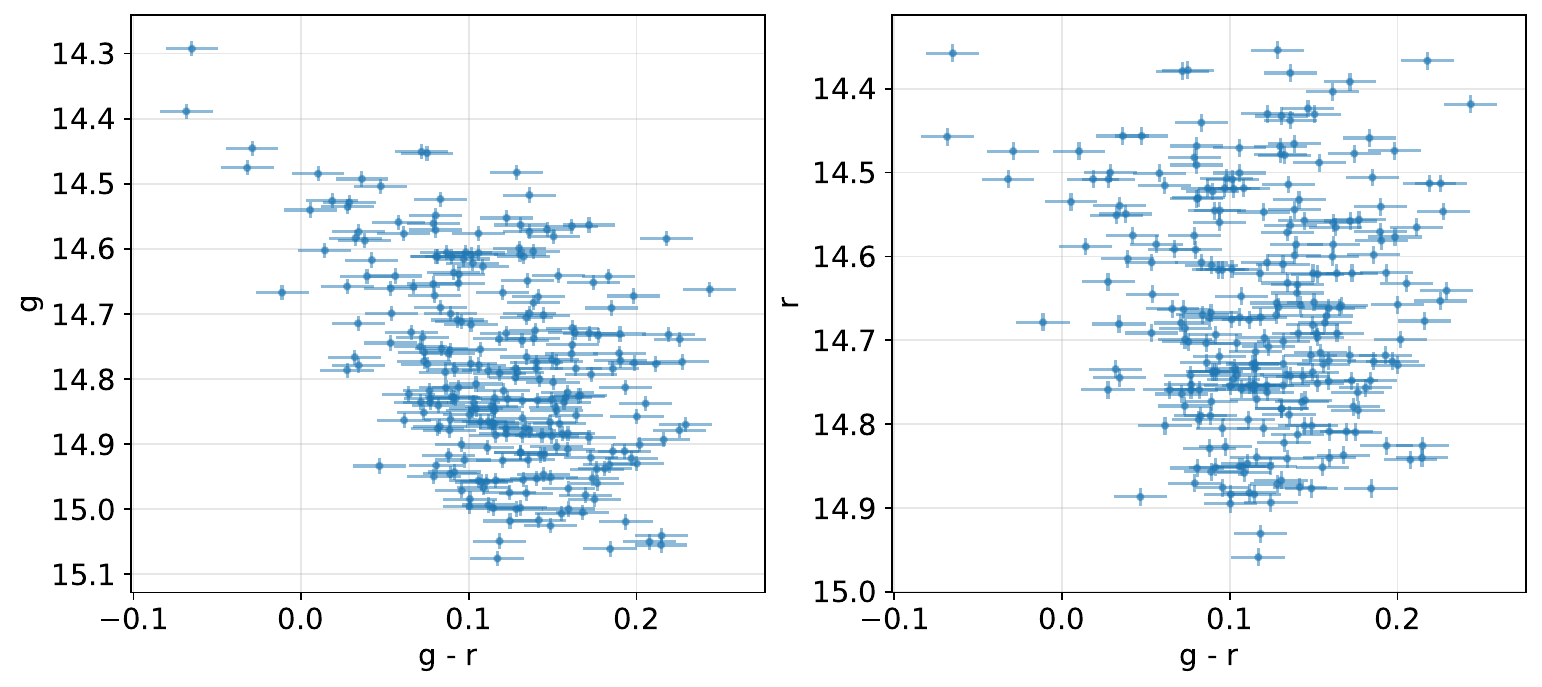}}}
\centerline{\resizebox*{1.0\columnwidth}{!}{\includegraphics[angle=0]{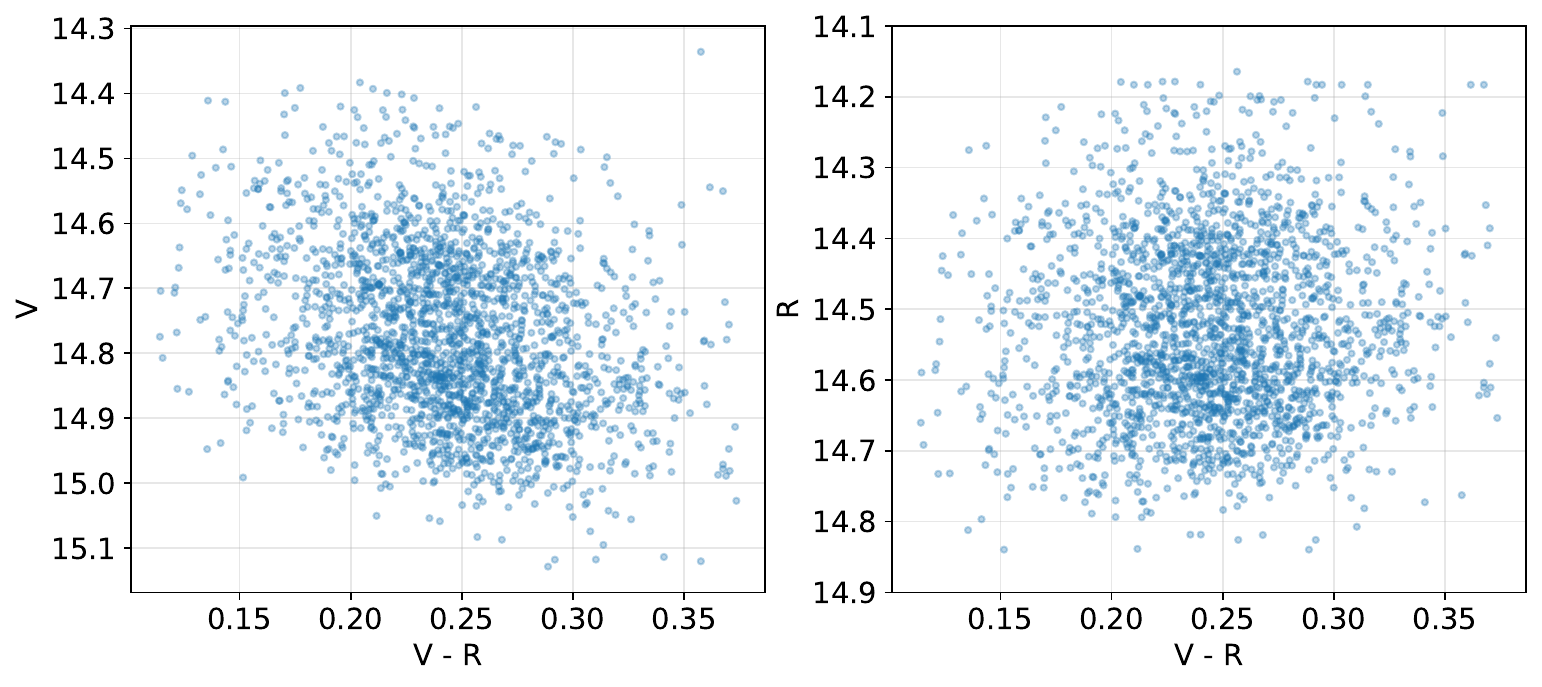}}}
\caption{Color-magnitude diagrams constructed from ZTF (upper panels) and FRAM-ORM (lower panels) light curves. Slight trends are noticeable in the bluest ($g$ and $V$) bands, while reddest ($r$ and $R$) are uncorrelated.}
\label{fig:color}
\end{figure}
In order to better constrain the short time scale variability of the object color, we also initiated a series of dedicated observations of \Jstar\, on FRAM-ORM robotic telescope \citep{FRAM2019}, which is a 10-inch Meade f/6.3 Schmidt-Cassegrain telescope with custom Moravian Instruments G2-1600 CCD installed in the Roque de Los Muchachos Observatory, La Palma.
The observations were performed since July till October 2022. The data were acquired in several sets every night, with several 120~s exposures in Johnson-Cousins $V$ and $R$ filters in every set, and then automatically processed by a dedicated Python pipeline based on the {\sc STDPipe} package \citep{stdpipe}, which includes bias and dark current subtraction, flat-fielding, cosmic ray removal, astrometric calibration, aperture photometry, and photometric calibration using the catalogue of synthetic photometry based on Gaia DR3 low-resolution BP/RP spectra \citep{gaiadr3syn}. The resulting light curve is shown in Figure~\ref{fig:lc_fram}. Its behaviour is essentially the same as in ZTF data, with irregular variability on the time scale of days to weeks and no systematic color evolution. Intra-night light curves shown in Figure~\ref{fig:fram_intranight} also show synchronous behaviour in both photometric bands, with variability apparent on time scale as short as tens of minutes.

While the light curve is clearly variable in all data sets we collected, we were unable to find any clear periodicity in it. The autocorrelation functions of the data from different instruments computed using a method of \citet{sacf} suitable for non-uniform time series are shown in Figure~\ref{fig:autocorr}. Their shapes are slightly different due to different time intervals and temporal samplings used, but all clearly shows characteristic variability time scales of several tens of days.

We also constructed the color-magnitude diagrams using the data from ZTF and FRAM-ORM (see Figure~\ref{fig:color}). While not conclusive, they both show slight dependence of the bluest band ($g$ for ZTF and $V$ for FRAM-ORM) on the color, while reddest band is not correlated with it. Such behaviour suggests that the amplitude of variability is slightly larger in bluest bands, which is consistent with the root mean square values for corresponding light curves shown in Figure~\ref{fig:lc_fram}. 

\begin{figure*}
\centerline{\resizebox*{2.0\columnwidth}{!}{\includegraphics[angle=0]{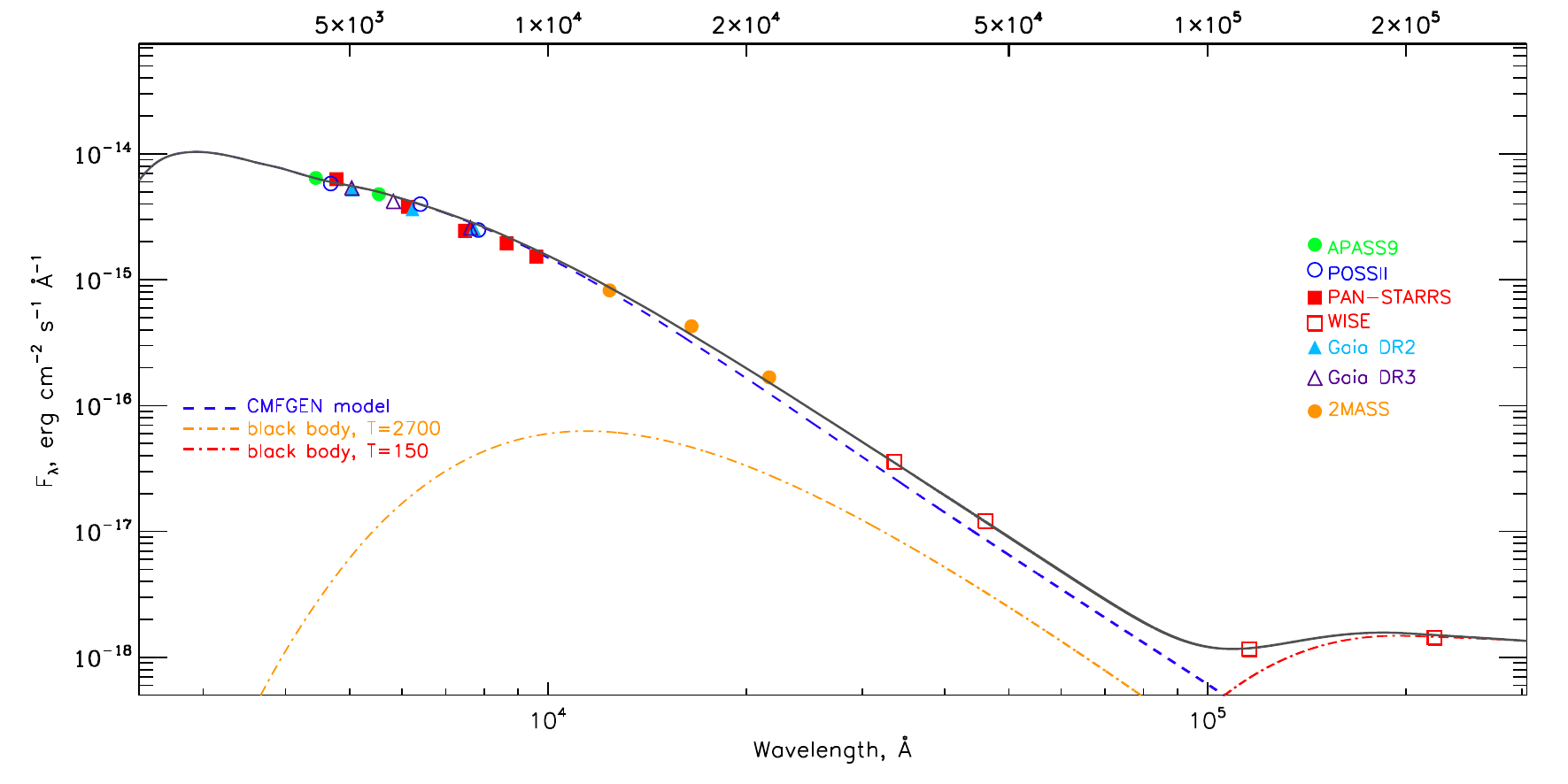}}}
\caption{Comparison of observed spectral energy distribution of \Jstar  based on the multi-wavelength photometric data with the model. The model includes reddened {\sc cmfgen} model spectrum (blue dashed line) and two blackbody components with $T=2700$~K and $T=150$~K, shown by orange and red dashed-doted lines, correspondingly. The model CMFGEN spectrum is scaled for the distance d=2.5~kpc and the interstellar extinction ($E{(B-V)} = 0.6$) is applied to it.  Photometric data are from APASS DR9 \citep{APASS9}, POSS\,II \citep{gsc22}, Pan-STARRS DR1 \citep{PanSTARRS2016}, WISE \citep{allWiseCutri}, 2MASS \citep{Cutri2MASS}, Gaia DR2 \citep{GaiaDR2} and DR3 \citep{GaiaEDR3} catalogues. }
\label{fig:sed}
\end{figure*}
\section{Discussion}\label{sec:discussion}

\subsection{Location in Hertzsprung--Russell diagram and evolutionary status of J0409+323} 

We constructed the qualitative model for \Jstar\, atmosphere using {\sc cmfgen} non-LTE radiative transfer code \citep{Hillier5}. To determine the effective temperature $T_{\rm eff}$  we used intensities of \CIII\,$\lambda$4647 and \CIV\,$\lambda$5801, 5812 lines, as well as ones of \HeI and \HeII. The model spectrum correctly reproduces the set of lines seen in the data, with the intensities close to observed ones, as well as overall spectral energy distribution (SED)  (Figure~\ref{fig:sed}). However, as we do not know the exact wind parameters, its terminal velocity, and chemical composition (such as He/H ratio and abundances of CNO elements) on such advanced evolutionary stage, we did not try to fit the exact profiles of individual spectral lines, and only estimated temperature of the star ($T_{\rm eff}=37000-41000$~K) and its luminosity ($L_*=900-1000~\Lsun$).  

By comparing an overall shape of the model spectrum with the photometric data acquired from several sky surveys (see Figure~\ref{fig:sed}) in optical and near-infrared ranges we also estimated the reddening towards \Jstar\, to be $E(B-V)$=0.6. It is 2.6 times larger than the total Galactic reddening in this direction (see Table~\ref{tab:par}). It suggests that additional extinction occurs in the circumstellar material around \Jstar. The presence of such material is also consistent with the infrared excess apparent in the SED as shown in Figure~\ref{fig:sed}. To describe this infrared excess we added two blackbody components with $T=2700$~K and $T=150$~K on top of the {\sc cmfgen} model. Thus we conclude that  \Jstar\, is surrounded by dust envelopes, both warm and cold, likely formed due to past eruptive activity. 

\begin{figure}
\centerline{\resizebox*{1.0\columnwidth}{!}{\includegraphics[angle=0]{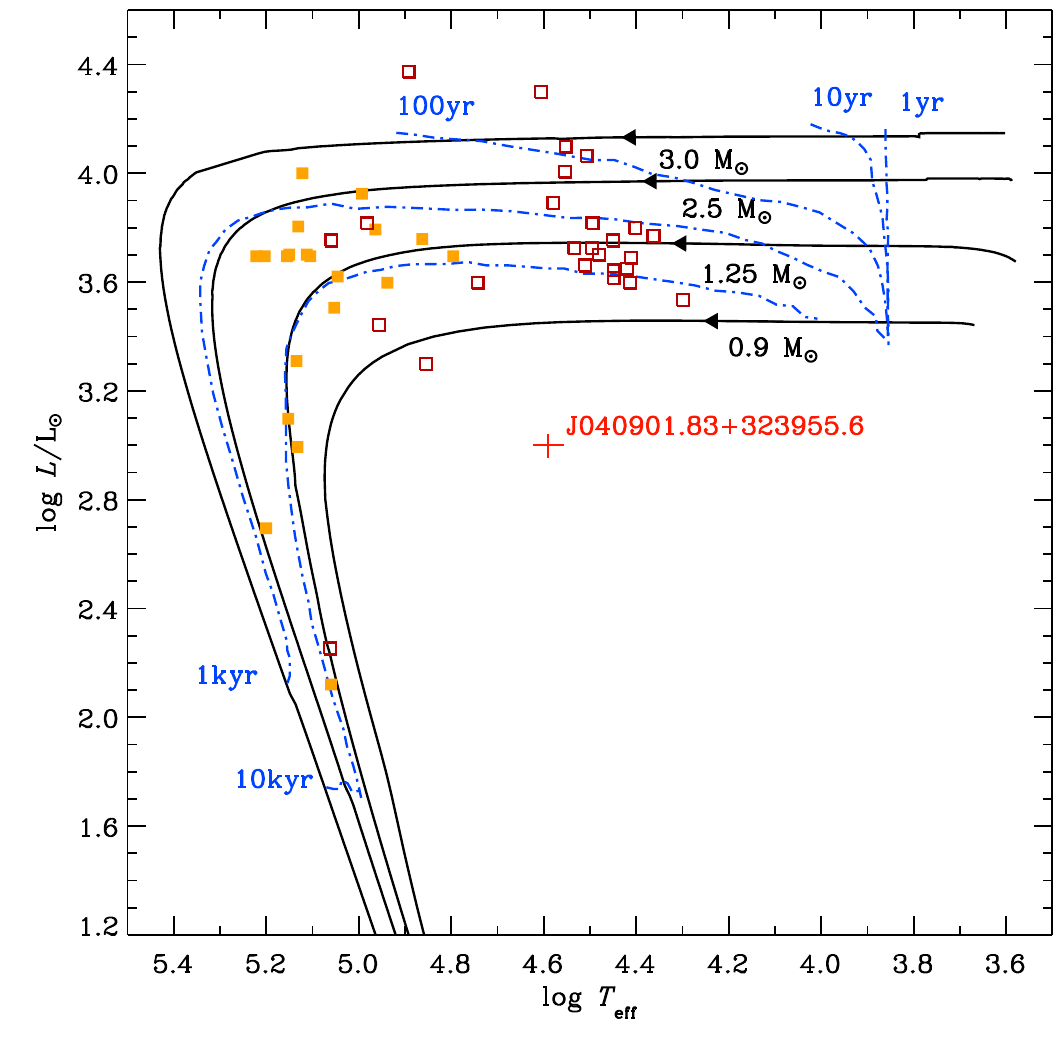}}}
\caption{Location of \Jstar\, in HR diagram. Filled orange squares mark the positions of known [WO] CSPNe, red ones -- [WC] CSPNe, taken from \citet{Weidmann2020}. Black solid lines show evolutionary tracks for objects of three different initial masses on H-burning post-AGB stage, while blue dash-dotted lines are corresponding isochrones for different ages since the moment of leaving the AGB, defined as
log $T_{\rm eff}=3.85$. Evolutionary tracks and isochrones are from \citet{Bertolami2016}. }
\label{fig:hrd}
\end{figure}


Figure~\ref{fig:hrd} shows  the location of \Jstar\ in HR diagram and evolutionary tracks for low mass stars on post-AGB stage \citep{Bertolami2016}. Numerical simulations of a post-AGB evolution show that stars become significantly hotter during approximately 1000 years after leaving the asymptotic giant branch, and rapidly move leftward in HR diagram \citep{Bertolami2016} before finally settling down as slowly cooling CSPNe. Low luminosity of  \Jstar\, $L_*\approx1000~\Lsun$ suggests that \Jstar\, is a very low mass star, near the limit for CSPN formation during the lifetime of the Galaxy. Indeed, the stars with initial mass less than $0.88~\Msun$ did not yet have enough time to leave the main sequence and evolve towards CSPNe \citep{Saracino2016,Weidmann2020}.

The spectrum of \Jstar displays prominent hydrogen lines (see Section~\ref{sec:spectroscopy}). They are typical for late-type WN stars \citep{Hamann2006,Crowther2007}, and thus were in agreement with the star' classification of \citet{Skoda2020}. On the other hand, they are not consistent with our new classification of \Jstar as [WC8-9]. However, the two-peaked shape of H$\alpha$  with 5\AA\ separation  
suggests that these lines are formed in the hydrogen-rich circumstellar shell surrounding the star that had already been ejected but did not yet expand significantly to be observable as a protoplanetary nebula like e.g. in \citet{GomezGonzalez2022}.  Shell expansion velocity derived from peak separation is about 120~\kms, which is significantly higher than the ones seen in nebular lines in planetary nebulae (20-30~\kms), suggesting that the shell is still much smaller and denser. Therefore, we interpret the presence of hydrogen lines and absence  of nebular lines in the spectrum of \Jstar as one more evidence that this object is in transition into [WR] stage right now. Indeed, it is already hot enough to be \textit{bona fide} CSPN, but its spectrum still lacks nebular emission lines. Also, no signs of the nebula around it are seen in direct images in IR range, while the SED shows warm and cold thermal components most probably corresponding to the dust surrounding the star.

\begin{figure}
\centerline{\resizebox*{1.0\columnwidth}{!}{\includegraphics[angle=0]{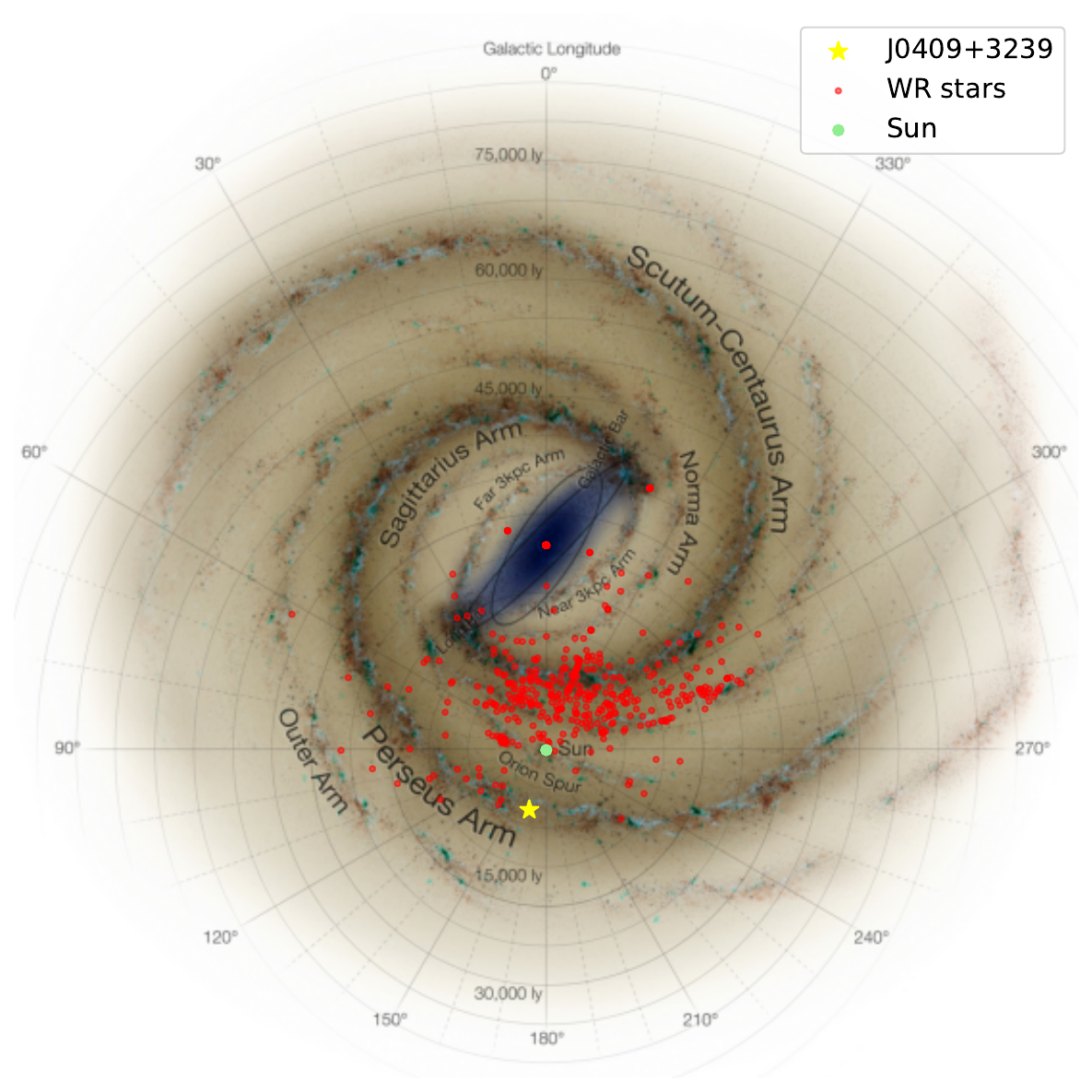}}}
\caption{The distribution of known WR stars from Galactic Wolf-Rayet Catalogue (\url{http://pacrowther.staff.shef.ac.uk/WRcat/index.php}) in the Milky Way, and the position of \Jstar. The distances are taken from \citet{GaiaEDR3}.}
\label{fig:galaxy}
\end{figure}
\begin{figure}
\centerline{\resizebox*{1.0\columnwidth}{!}{\includegraphics[angle=0]{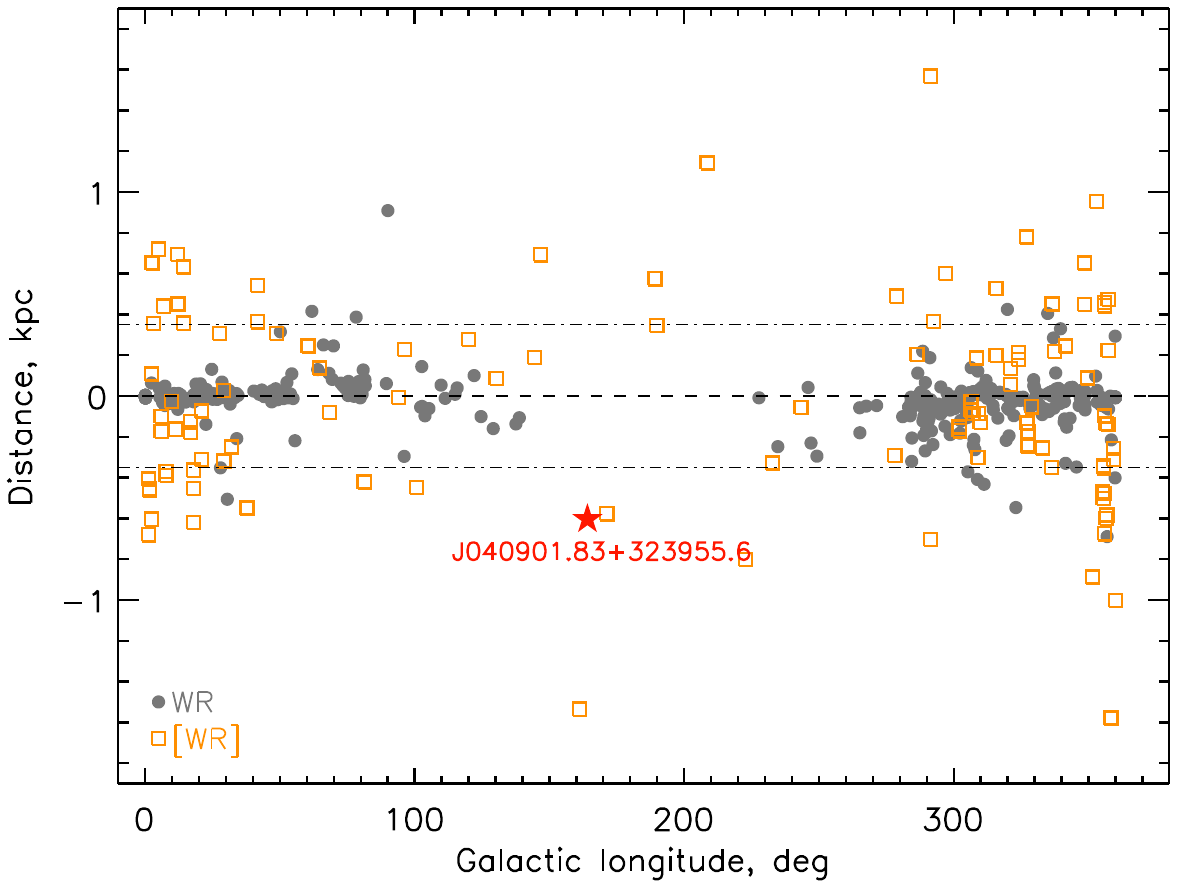}}}
\caption{Distribution WRs and [WRs] from galactic plane. Orange squares are Galactic [WR] stars (list taken from \citep{Weidmann2020}) distances for them taken from DR3, while grey circles -- WR stars, according \href{http://pacrowther.staff.shef.ac.uk/WRcat/index.php}{Galactic Wolf-Rayet Catalogue}  -- distances for them taken from same database. Dashed line shows centre of Galactic plane, while dashed-doteed lines mark borders of the think disc at 0.35~kpc. }
\label{fig:altitude}
\end{figure}

\subsection{Position in the Galaxy}\label{sec:galaxy}

Our initial hypothesis about \Jstar\, being a low mass object with WR phenomenon was based on the distance estimation. As mentioned before, according to {\it Gaia} DR3 the distance to \Jstar\, is 2.5~kpc, which means that the object lies in the Perseus Galactic arm (Figure~\ref{fig:galaxy}). There is no reason to believe that this distance estimation is wrong -- indeed, while there is another spiral arm, an Outer Arm, in that direction, moving \Jstar\, there would only increase its luminosity by approximately 4 times, which is not enough to make it classical WR star.
It is interesting to note that  \Jstar\, is located just 21 degrees from Galactic anti-center direction, in a sector between $138.9\degr$ - $227.8\degr$  Galactic longitude where no WR were ever found\footnote{It was first stated by \citet{Roberts1962}, and it is still true \citep{van_der_Hucht2003, Rustamov2023}, see also \url{http://pacrowther.staff.shef.ac.uk/WRcat/index.php}.}, which also indirectly supports that \Jstar\, is a   low mass object with WR phenomenon.

We considered the distance of \Jstar\, from the Galactic plane and compared its location with the distribution of Galactic WR and [WR] stars (Figure~\ref{fig:altitude}). While $\sim50\%$ of [WR]s are located above the Galactic plane (farther than ${\pm}$0.35~kpc), only $\sim3\%$ of WRs are located outside of the thin Galactic disk.  For \Jstar\, the distance above Galactic plane is -0.725~kpc, i.e. the star is a member of Galactic halo.

\subsection{Variability} 

Analysis of photometric data in Section~\ref{sec:photometry} clearly shows that \Jstar\, 
displays significant aperiodic variability on time scales from hours till tens of days, with amplitude slightly larger in bluer bands. We had not been able to find similar photometric behaviour being mentioned for other known [WR] stars in the literature. Thus, we checked the variability of CSPN stars classified as [WR] from the list of \citet{Weidmann2020} in ZTF data archive. We found the light curves having more than 100 photometric measurements for 19 stars including V348\,Sgr, the variable classified as R\,Coronae Borealis (RCB) type. The latter is the only significantly variable object in the sample -- the rest does not show any signs of a systematic variability comparable in amplitude or pattern to the one we see for \Jstar.  On the other hand, the variability with similar amplitudes on similar time scales and with similar color-magnitude behaviour was previously detected in carbon-rich protoplanetary nebulae \citep{Hrivnak2022}, where it is attributed to the opacity variations of the circumstellar dust.

Spectral variability of  CSPNe stars usually manifests in changes of emission line intensities and changes of P Cygni profiles. It was detected in both ultraviolet  \citep{Feibelman1992, Patriarchi1995, Guerrero2013} and optical \citep{Arkhipova2013,Kondratyeva2017} ranges. Unfortunately, for \Jstar\, no UV data are available, and the only variation in the optical spectra separated by 8 years is the change in the intensity of \CIII~$\lambda$5695.92 line.  
Thus, we cannot yet relate it to either observed photometric variability or the evolutionary changes on the time scales of 8 years between the spectra, and additional spectral monitoring is definitely required.  In general, we suggest that this peculiar variability, both photometric and spectral, confirms the transitional evolutionary stage of \Jstar. 
\section{Conclusion}\label{sec:conclusion}

We performed detailed study of \Jstar, originally selected from LAMOST spectral archive and classified as a WN by \citet{Skoda2020}. Based on its luminosity, location in the Galaxy and position in color-magnitude and HR diagrams
we speculate that \Jstar is on a transitional post-AGB stage evolving towards CSPN, with actually spectral type [WC8-9]. The object displays peculiar pattern of aperiodic variability with characteristic time scales from hours to tens of days which is not observed in other known CSPN, which we attribute to its earlier evolutionary stage. The  absence of forbidden lines, presence of warm and cool thermal components in the SED, and double-peaked H$\alpha$ also suggest that \Jstar very recently made transition between post-AGB and CSPN, did not yet develop a nebula, but is already surrounded by a shell-like structure which is typical for the protoplanetary nebulae. The very low initial mass of the object (less than $0.9~\Msun$) is on the limit of CSPN formation during the life time of the Galaxy.

\section*{Acknowledgements}

We are grateful to Walter Weidmann (the referee) for useful comments
and suggestions on the manuscript, also to Helge Todt and  Marcelo Miguel Miller Bertolami for helpful discussion during the development of this work. 
The work is funded from the European Union's Framework Programme for Research and Innovation Horizon 2020 (2014-2020) under the Marie Sk\l{}odowska-Curie Grant Agreement No. 823734. 
O.M. acknowledges ﬁnancial support by the Astronomical Institute of the Czech Academy of Sciences through the project RVO:67985815.
S.K. acknowledges support from the European Structural and Investment Fund
and the Czech Ministry of Education, Youth and Sports (Project CoGraDS --
CZ.02.1.01/0.0/0.0/15\_003/0000437).

Guoshoujing Telescope (the Large Sky Area Multi-Object Fiber Spectroscopic Telescope LAMOST) is a National Major Scientific Project built by the Chinese Academy of Sciences. Funding for the project has been provided by the National Development and Reform Commission. LAMOST is operated and managed by the National Astronomical Observatories, Chinese Academy of Sciences.

This study is partially based on the data obtained at the  unique scientific facility   the Big Telescope Alt-azimuthal of SAO RAS and  was supported  under  the   Ministry of Science and Higher Education of the Russian Federation grant  075-15-2022-262 (13.MNPMU.21.0003). 

The operation of the robotic telescope FRAM-ORM is supported by the grants of the Ministry of Education of the Czech Republic LM2023032 and LM2023047. 
The data calibration and analysis related to the FRAM-ORM telescope are supported by the Ministry of Education of the Czech Republic EU/MEYS funds CZ.02.1.01/0.0/0.0/16\_013/0001402, CZ.02.1.01/0.0/0.0/16\_019/0000754, CZ.02.1.01/0.0/0.0/18\_046/0016010 and CZ.02.01.01/00/22\_008/0004632.

This work is based in part on observations obtained with the Samuel Oschin 48-inch Telescope at the Palomar Observatory as part of the Zwicky Transient Facility (ZTF) project. ZTF is supported by the National Science Foundation under Grant No. AST-1440341 and a collaboration including Caltech, IPAC, the Weizmann Institute for Science, the Oskar Klein Center at Stockholm University, the University of Maryland, the University of Washington, Deutsches Elektronen-Synchrotron and Humboldt University, Los Alamos National Laboratories, the TANGO Consortium of Taiwan, the University of Wisconsin at Milwaukee, and Lawrence Berkeley National Laboratories. Operations are conducted by Caltech Optical Observatories (COO), the Infrared Processing and Analysis Center (IPAC), and the University of Washington (UW).

This work has made use of data from the Asteroid Terrestrial-impact Last Alert System (ATLAS) project. The Asteroid Terrestrial-impact Last Alert System (ATLAS) project is primarily funded to search for near earth asteroids through NASA grants NN12AR55G, 80NSSC18K0284, and 80NSSC18K1575; byproducts of the NEO search include images and catalogs from the survey area. This work was partially funded by Kepler/K2 grant J1944/80NSSC19K0112 and HST GO-15889, and STFC grants ST/T000198/1 and ST/S006109/1. The ATLAS science products have been made possible through the contributions of the University of Hawaii Institute for Astronomy, the Queen’s University Belfast, the Space Telescope Science Institute, the South African Astronomical Observatory, and The Millennium Institute of Astrophysics (MAS), Chile.


It also uses the data from the Pan-STARRS1 Surveys (PS1) and the PS1 public science archive, whose have been made possible through contributions by the Institute for Astronomy, the University of Hawaii, the Pan-STARRS Project Office, the Max-Planck Society and its participating institutes, the Max Planck Institute for Astronomy, Heidelberg and the Max Planck Institute for Extraterrestrial Physics, Garching, The Johns Hopkins University, Durham University, the University of Edinburgh, the Queen's University Belfast, the Harvard-Smithsonian Center for Astrophysics, the Las Cumbres Observatory Global Telescope Network Incorporated, the National Central University of Taiwan, the Space Telescope Science Institute, the National Aeronautics and Space Administration under Grant No. NNX08AR22G issued through the Planetary Science Division of the NASA Science Mission Directorate, the National Science Foundation Grant No. AST-1238877, the University of Maryland, Eotvos Lorand University (ELTE), the Los Alamos National Laboratory, and the Gordon and Betty Moore Foundation.


\section*{DATA AVAILABILITY}
The data underlying this article will be shared on reasonable request to the corresponding author.

\bibliographystyle{mnras}
\bibliography{New_WR.bib}

\bsp	
\label{lastpage}
\end{document}